\documentclass[aps,prl,preprint,groupedaddress,floatfix]{revtex4}

\usepackage{graphicx}% Include figure files
\usepackage{dcolumn} % Align table columns on decimal point
\usepackage{bm}      % bold math

\begin{document}

\title{Verification of universal relations in a strongly interacting Fermi gas}
\author{J. T. Stewart}
\author{J. P. Gaebler}
\email[Electronic address: ]{gaeblerj@jila.colorado.edu}
\author{T. E. Drake}
\author{D. S. Jin}
\homepage[URL: ]{http://jilawww.colorado.edu/~jin/}

\affiliation{JILA, Quantum Physics Division, National Institute of
Standards and Technology and Department of Physics, University of
Colorado, Boulder, CO 80309-0440, USA}

\date{\today} \begin{abstract}
Many-body fermion systems are important in many branches of physics,
including condensed matter, nuclear, and now cold atom physics. In many
cases, the interactions between fermions can be approximated by a
contact interaction. A recent theoretical advance in the study of
these systems is the derivation of a number of exact universal
relations that are predicted to be valid for all interaction
strengths, temperatures, and spin compositions
\cite{Tan08,Tan08a,Tan08b,Braaten08,Castin09,Zhang2009}. These
equations, referred to as the Tan relations, relate a microscopic quantity, namely, the amplitude of
the high-momentum tail of the fermion momentum distribution, to the
thermodynamics of the many-body system. In this work, we provide
experimental verification of the Tan relations in a strongly
interacting gas of fermionic atoms. Specifically, we measure the
fermion momentum distribution using two different techniques, as
well as the rf excitation spectrum and determine the effect of
interactions on these microscopic probes. We then measure the
potential energy and release energy of the trapped gas and test the
predicted universal relations.
\end{abstract}

 \pacs{??)}

\maketitle

%%Beginning of Text

In 2008, Shina Tan derived a number of universal relations for an
interacting Fermi gas with short-range, or contact, interactions
\cite{Tan08,Tan08a,Tan08b}.  These relations are very powerful
because they connect microscopic quantities, such as the momentum
distribution of the fermions, to macroscopic quantities, such as the
total energy of the system. Furthermore, the relations are universal
in that they do not depend on the details of the interparticle
potential that gives rise to the interaction, nor do they depend
on the state of the system, which could be an exotic Fermi
superfluid, a normal Fermi liquid, or even a simple two-body state
such as a diatomic molecule composed of two fermionic atoms. At the
heart of the universal relations is a single quantity, which Tan
termed the contact.  The contact is defined as the amplitude of
the high-$k$ tail of the momentum distribution $n(k)$, which was
previously predicted to scale as $1/k^4$ for an interacting Fermi
gas \cite{Viverit2004}.  Remarkably, it can be shown that the
contact encapsulates all of the many-body physics \cite{Zhang2009}.
Therefore, it is predicted that by measuring this tail of the
momentum distribution and then applying universal relations, one
could determine many other properties of the system.  Two recent papers report the contact for a
strongly interacting Fermi gas, extracted from photoassociation measurements
\cite{Partridge2005,Castin09} and inelastic Bragg spectroscopy
\cite{Vale2010} and compare the results with theoretical
predictions for the BCS-BEC crossover.  Here, we present a series of
measurements that not only measure the contact in the BCS-BEC
crossover with two different techniques, but moreover test the Tan
relations experimentally by comparing measurements of both
microscopic and macroscopic quantities in the same system. Our
results directly verify the universal relations by exploiting the
fact that while the value of the contact depends on the many-body
state and on parameters such as temperature, number density, and
interaction strength, the universal relations do not.

Our measurements are done in an ultra cold gas of fermionic $^{40}$K
atoms confined in a harmonic trapping potential.  We cool the gas to
quantum degeneracy in a far-detuned optical dipole trap as described
in previous work \cite{Stewart2008}. The trap is axially symmetric
and parameterized by a radial trap frequency, which varies for these
data from $\omega_{r} = 2\pi\cdot230$ to $2\pi\cdot260$ Hz and an
axial trap frequency, which varies from $\omega_z = 2\pi\cdot17$ to
$2\pi\cdot21$ Hz. We obtain a $50/50$ mixture of atoms in two spin
states, namely the $|f,m_f\rangle= |9/2, -9/2\rangle$ and
$|9/2,-7/2\rangle$ states, where $f$ is the total atomic spin and
$m_f$ is the projection along the magnetic-field axis. Our final
stage of evaporation occurs at a magnetic field of 203.5 G, where
the $s$-wave scattering length, $a$, that characterizes the
interactions between atoms in the $|9/2, -9/2\rangle$ and
$|9/2,-7/2\rangle$ states is approximately 800 $a_0$, where $a_0$ is
the Bohr radius. At the end of the evaporation, we have $10^5$ atoms
per spin state at a normalized temperature $\frac{T}{T_F} = 0.11\pm
0.02$ where the Fermi temperature corresponds to the Fermi Energy, $E_F = k_b T_F = \hbar \omega (6 N)^{1/3}$ where $N$ is the total
atom number in one spin state and $\omega = (\omega_r^2 \omega_z)^{1/3}$ and $k_b$ is the the Boltzmann constant.  After the evaporation we increase the interaction strength adiabatically
with a slow magnetic-field ramp to a Fano-Feshbach scattering
resonance.

The momentum distribution of the fermions, $n(k)$, is predicted to
scale as $1/k^4$ at high $k$, with the contact being the coefficient
of this high momentum tail.  Following Tan \cite{Tan08a}, we define
the integrated contact per particle for the trapped gas, which we will refer to
simply as the contact, using
\begin{equation}\label{eq:Contact} C = \lim_{k \rightarrow
\infty} k^4 n(k)  .
\end{equation}
Here, $k$ is the wave number in units of the Fermi wave number,
$k_F= \frac{2 m E_F}{\hbar}$, and $n(k)$ for a 50/50 spin mixture is
normalized such that
%\begin{equation}\label{eq:momdist}
$\int_{0}^{\infty}\frac{n(k)}{(2 \pi)^3}d^3 k = 0.5.$
%\end{equation}
Note that the contact is expected to be the same for both spin
states in the interacting Fermi gas, even in the case of an
imbalanced spin mixture.  Theoretically, the contact is defined in the limit of
$1<<k<<1/(k_Fr_0)$, where $r_0$ is the range of the interatomic
potential. Using a typical value of $k_F=\frac{1}{2200} a_0^{-1}$ for our trapped
$^{40}$K gas and the van der Waals length, $r_0 = 60 a_0$, we find
$1/(k_Fr_0)=37$.

We directly measure $n(k)$ using ballistic expansion of the trapped
gas, where we turn off the interactions for the expansion.  We
accomplish this by rapidly sweeping the magnetic field to $209.2$ G
where $a$ vanishes, and then immediately turning off the external
trapping potential \cite{Regal2005c}.  We let the gas expand for $6$
ms before taking an absorption image of the cloud. The probe light
for the imaging propagates along the axial direction of the trap and
thus we measure the radial momentum distribution. Assuming the
momentum distribution is spherically symmetric, we obtain the full
momentum distribution with an inverse Abel transform.

\begin{figure}
\includegraphics[width=14 cm]{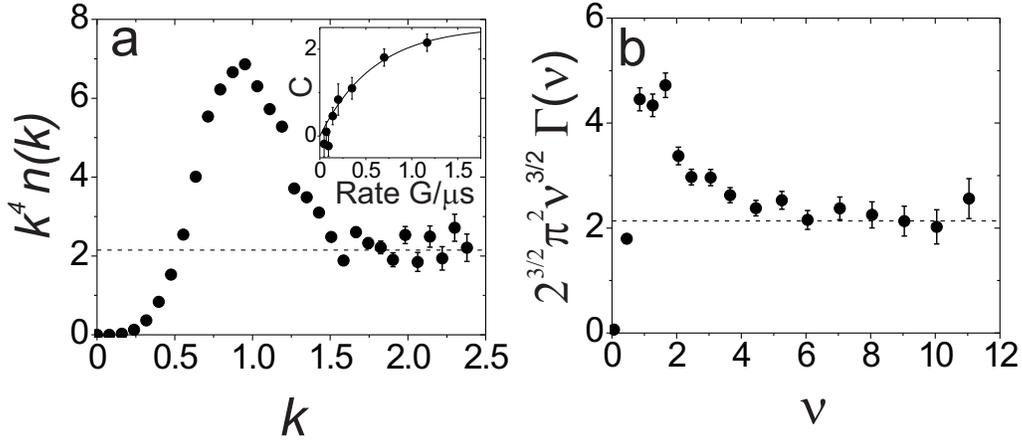}
\caption{\textbf{Extracting the contact from the momentum
distribution and rf lineshape.} (a) Measured momentum distribution
for a Fermi gas at $\frac{1}{k_F a}= -0.08 \pm0.10$. Here, the wave
number $k$ is given in units of $k_F$, and we plot the normalized
$n(k)$ multiplied by $k^4$.  The dashed line corresponds to
$C=2.15\pm 0.2$, which is obtained by averaging $k^4n(k)$ for $k
> 1.85$. (Inset) The measured value for $C$ depends on the
rate of the magnetic-field ramp that turns off the interactions
before time-of-flight expansion. (b) RF lineshape measured for a
Fermi gas at $\frac{1}{k_F a}=-0.03 \pm0.10$. Here, $\nu$ is the
rf detuning from the single-particle Zeeman resonance, given in
units of $E_F/h$. We plot the normalized rf lineshape multiplied by
$2^{3/2} \pi^2 \nu^{3/2} $, which is predicted to
asymptote to $C$ for large $\nu$.  Here, we obtain $C=2.13\pm 0.2$
(dashed line) from an average of the data for $\nu
> 5$.} \label{figure1}
\end{figure}

Fig. \ref{figure1}a shows an example $n(k)$ for a strongly interacting gas
measured with this technique.    For this data, the dimensionless interactions strength $(k_F a)^{-1}$, is $-0.08 \pm 0.10$.  Empirically, we find that the measured $n(k)$
exhibits a $1/k^4$ tail and we extract the contact $C$
from the average value of $k^4n(k)$ for $k>k_C$ where we use $k_C = 1.85$ for $(k_F a)^{-1} > -0.5$, and $k_C = 1.55$ for  $(k_F a)^{-1} < -0.5$. One issue is whether or not
the interactions are switched off sufficiently quickly to accurately
measure the high-$k$ part of the $n(k)$. The data in Fig.
\ref{figure1}a were taken using a magnetic-field sweep rate of
approximately $1.4 \frac{G}{\mu s}$ to turn off the interactions for
the expansion.  In the inset to Fig. \ref{figure1}a, we show the
dependence of the measured $C$ on the magnetic-field sweep rate.
Using an empirical exponential fit (line in Fig. \ref{figure1}a
inset), we estimate that at our typical sweep rate of approximately
$1.4 \frac{G}{\mu s}$ our measured $C$ is systematically low by
about $10\%$.  We have therefore scaled the contact measured with this method by $1.1$.

The contact is also manifest in rf spectroscopy, where one applies a
pulsed rf field and counts the number of atoms that are transferred
from one of the two original spin states into a third, previously
unoccupied, spin state \cite{Regal2003a}.  We transfer atoms from the $|9/2,-7/2\rangle$ state to the $|9/2,-5/2\rangle$ state. It is predicted that the number of atoms
transferred as a function of the rf frequency, $\nu$, scales as
$\nu^{-3/2}$ for large $\nu$, and that the amplitude of
this high frequency tail is $\frac{C}{2^{3/2}\pi^2}$
\cite{Pieri2009,Schneider2009,Braaten10}. Here, $\nu=0$ is the
single-particle spin-flip resonance, and $\nu$ is given in
units of $E_F/h$. This prediction requires that atoms transferred to the third spin-state have only weak interactions with the other atoms, so that ``final state effects" are negligible \cite{Braaten10,Chin2005,Yu2006,Punk2007,Basu2007,Perali2008,Veillette2008,He2008}, as is the case for $^{40}K$ atoms. In Fig. \ref{figure1}b, we plot a measured rf spectrum multiplied by
$2^{3/2}\pi^2\nu^{3/2} $. The rf spectrum, $\Gamma(\nu)$, is
normalized so that the integral over the rf lineshape equals $0.5$.
Empirically, we observe the predicted $1/\nu^{3/2}$ behavior
for $\nu>\nu_{C}$. To obtain the contact we average $2^{3/2}\pi^2\nu^{3/2}\Gamma(\nu)$ for $\nu > \nu_C$ where $\nu_{C} = 5$ for $
(k_F a)^{-1} > -0.5$, and $\nu_C = 3$ for  $(k_F a)^{-1} < -0.5$.

The connection between the tail of the rf spectrum and the high-$k$
tail of the momentum distribution can be seen in the Fermi spectral
function, which can be probed using photoemission spectroscopy for
ultra cold atoms \cite{Stewart2008}.  Recent photoemission
spectroscopy results on a strongly interacting Fermi gas \cite{contactnote1}
revealed a weak, negatively dispersing feature at high $k$ that
persists to temperatures well above $T_F$. This feature was
attributed to the effect of interactions, or the contact, consistent
with a recent prediction that the $1/k^4$ tail in $n(k)$ should
correspond to a high-$k$ part of the spectral function that
disperses as $-k^2$ \cite{Schneider2009b}.  Atom photoemission
spectroscopy, which is based upon momentum-resolved rf spectroscopy,
also provides a method for measuring $n(k)$. By integrating over
the energy axis, or equivalently, summing data taken for different
rf frequencies, we obtain $n(k)$.  This alternative method for
measuring $n(k)$ yields results similar to the ballistic expansion
technique, but avoids the issue of magnetic-field ramp rates.

\begin{figure}
\includegraphics[width=10 cm]{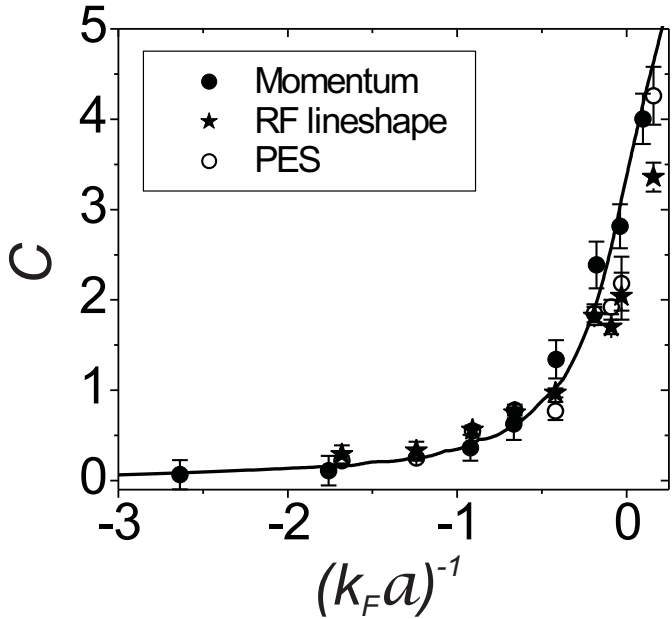}
\caption{\textbf{The contact.} We measure the contact, $C$, as a function of $(k_F a)^{-1}$ using three
different methods.  Filled circles correspond to direct measurements
of the fermion momentum distribution $n(k)$ using a fast
magnetic-field sweep to project the many-body state onto a
non-interacting state. The momentum distribution can then be
measured in ballistic expansion. Open circles correspond to $n(k)$
obtained using atom photoemission spectroscopy measurements.  Stars
correspond to the contact obtained from rf spectroscopy.  The
values obtained with these different methods show good agreement.  The
contact is nearly zero for a weakly interacting Fermi gas with
attractive interactions (left hand side of plot) and then increases as the
interaction strength increases to the unitarity regime where
$(k_F a)^{-1}=0$. The line is a theory curve obtained from Ref. \cite{Castin09}}. \label{figure2}
\end{figure}

In Fig. \ref{figure2} we show the measured contact for different
values of the dimensionless interaction strength, $1/k_Fa$.  Here,
the contact is extracted using the three different techniques
described above to probe two distinct microscopic quantities, namely
the momentum distribution and the rf lineshape. We find that the
amplitude of the $1/k^4$ tail of $n(k)$ and the coefficient of the
$1/\nu^{3/2}$ tail of the rf spectra yield consistent values
for $C$.  The solid line is a prediction for the contact that was reported in Fig. 1 of Ref. \cite{Castin09}.  This prediction consists of the BCS limit, interpolation of Monte Carlo data near unitarity, and the BEC limit for a trapped gas at zero temperature and uses a local density approximation.

Remarkably, the Tan relations predict that the contact, as revealed
in probes of the microscopic behavior of the gas, is directly
connected to the thermodynamics of the gas. To test the Tan
relations, we measure the potential energy and release energy of the
cloud. The total energy of the trapped gas divided by the number of
particles, $E$, is the sum of three contributions, the kinetic energy
$T$, the external potential energy $V$, and the interaction energy
$I$. Because the momentum distribution has a $1/k^4$ tail, the
kinetic energy obtained by integrating over the momentum
distribution diverges. However, the release energy $T+I$, which can
be measured in the usual time-of-flight expansion without turning
off the interactions, is not divergent.

 We measure $V$ by imaging the spatial
distribution of the atom cloud, similar to measurements that were
performed in Ref. \cite{Stewart2006}. We allow the cloud to expand
for $1.6$ ms to lower the optical density and then image along one
of the radial directions in order to see the density distribution in
the axial direction.  Because the expansion time is short compared
to the axial trap period (40 times shorter) the density distribution
in the axial direction reflects the in-trap axial density
distribution.  The potential energy per particle, in units of $E_F$,
is then $V = \frac{3}{N E_F} \frac{1}{2} m \omega_z^2 \langle z^2 \rangle$, where
$\langle z^2 \rangle$ is the mean squared width of the cloud in the axial
direction and we have assumed that the total potential energy is
distributed equally over the three axes.

\begin{figure}
\includegraphics[width=10 cm]{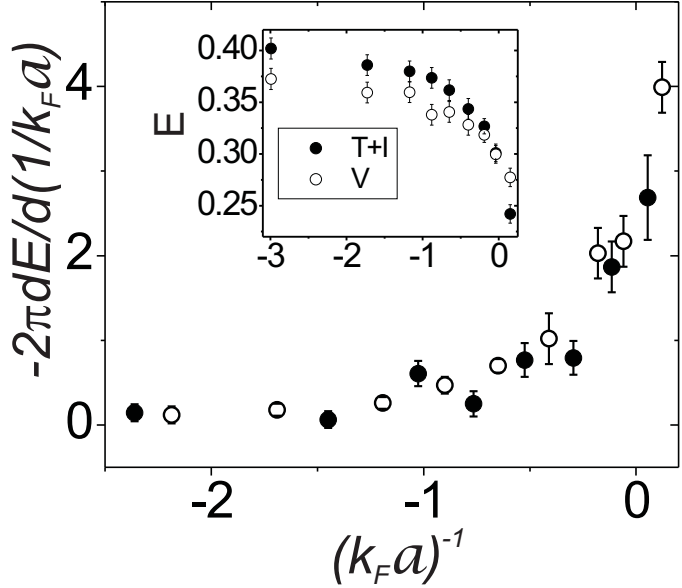}
\caption{\textbf{Testing the adiabatic sweep theorem.} (Inset) The
measured potential energy, $V$, and release energy, $T+I$, per particle in units of $E_F$ is shown as a function of
$1/k_Fa$.  (Main) Taking a discrete derivative of the data shown in
the inset, we find that $2\pi\frac{dE}{d(-1/(k_Fa))}$ ($\bullet$)
agrees well with the contact $C$ measured from the high-$k$ tail
momentum distribution ($\circ$).} \label{figure3}
\end{figure}

To measure the release energy $T+I$ we turn off the trap suddenly
and let the cloud expand for $t = 16$ ms (with interactions) before
imaging along one of the radial directions; this is similar to
measurements reported in Ref. \cite{Bourdel2003a}. The total release
energy is the sum of the release energy in the two radial directions
and the release energy in the axial direction.  For the radial
direction, the release energy per particle, in units of $E_F$, is simply $ T_{r}+ I_{r} = \frac{2}{N
E_F} \frac{1}{2} m \frac{\langle y^2\rangle }{t^2}$ where t is the expansion time
and $\langle y^2\rangle $ is the mean squared width of the expanded cloud in the
radial direction. For the axial direction, the expansion is slower
and the expanded cloud may not be much larger than the in-trap
density distribution. This is especially true near the Feshbach
resonance where the cloud expands hydrodynamically
\cite{O'Hara2002a}. Accounting for this,
the axial release energy is $T_z+I_z = \frac{1}{N E_F} \frac{1}{2} m
\frac{\langle z^2\rangle -z_0^2}{t^2}$, where $z_0^2$ is mean squared axial width
of the in-trap density distribution. We extract the mean squared
cloud widths from surface fits to the images, where we fit to a
finite temperature Fermi Dirac distribution while minimizing the
difference in energy between the raw data and the fit. Rather than
being theoretically motivated, we simply find empirically that this
functional form fits well to our images.  To eliminate systematic
error due to uncertainty in the trap frequencies and imaging
magnification, we measure the release energy and potential energy of
a very weakly interacting Fermi gas at $\frac{T}{T_F} = 0.11$, where
$T+I$ and $V$ for an ideal Fermi gas is $0.40 E_F$. We then use the
ratio of $0.40 E_F$ to our measured values as a multiplicative
correction factor that we apply to all of the data. This correction is within $5\%$ of unity.  For the point with $\frac{1}{k_F a}> 0$ we take into account the binding energy of the molecules in our tabulation of the release energy $T+I$ by adding $-1/(k_Fa)^2$. We show our data for the $V$ and $T+I$ versus $(k_F a)^{-1}$ in the inset of Fig. \ref{figure3}.

\begin{figure}
\includegraphics[width=10 cm]{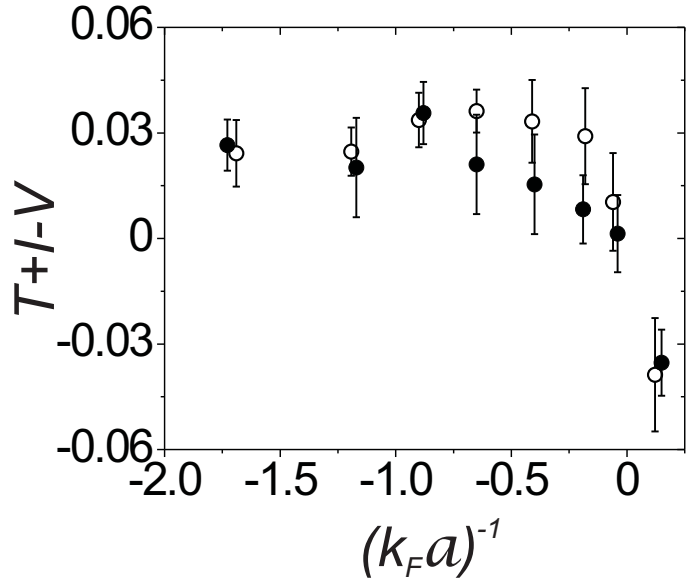}
\caption{\textbf{Testing the generalized virial theorem.} The difference between the measured release energy and potential energy per particle $T+I-V$ is shown as open circles. This corresponds to the left hand side of Eq. \ref{eq:GenVirTheorem}.
Filled circles show the right hand side of Eq.
\ref{eq:GenVirTheorem} obtained from the average values of the
contact shown in Fig. \ref{figure2}. The two quantities are equal
to within the measurement uncertainty, which is on order of $0.01
E_F$.} \label{figure4}
\end{figure}

We can now test the predicted universal relations connecting the
$1/k^4$ tail of the momentum distribution with the thermodynamics of
the trapped Fermi gas.  We first consider the adiabatic sweep
theorem \cite{Tan08},
\begin{equation}\label{eq:AdiabaticSweep}
2 \pi \frac{dE}{d(-1/(k_Fa))}=C,
\end{equation}
which relates the contact $C$ to the change in the total energy of
the system when the interaction strength is changed adiabatically. To obtain the energy per particle, $E$, we sum the values for $T+I$ and $V$ shown in the inset of Fig. \ref{figure3}.  To test the adiabatic sweep
theorem, we find the derivative, $\frac{dE}{d(-1/(k_Fa))}$, simply
by calculating the slope for pairs of neighboring points in the
inset to Fig. \ref{figure3}.  In the main part of Fig. \ref{figure4},
we compare this point-by-point derivative, multiplied by $2\pi$,
to $C$ obtained from the average values of
the data shown in Fig. \ref{figure2}($\circ$).  Comparing these
measurements of the left and right sides of Eqn.
\ref{eq:AdiabaticSweep}, we find good agreement and thus verify the
adiabatic sweep theorem for our strongly interacting Fermi gas.

A second universal relation that we can directly test is the
generalized virial theorem \cite{Tan08a},
\begin{equation}\label{eq:GenVirTheorem}
E-2V=T+I- V = -\frac{C}{4 \pi k_F a},
\end{equation}
which relates the difference between the release energy and the
potential energy to the contact.  Eqn. \ref{eq:GenVirTheorem} is
predicted to be valid for all values of the interaction strength
$(k_Fa)^{-1}$.  This generalized virial theorem reduces to $E-2V=0$ for
the ideal gas, where $I=0$, as well as for the unitarity gas, where
$(k_Fa)^{-1}=0$. This result for the unitarity gas was previously
verified in Ref. \cite{Thomas05}. Here, we test Eqn.
\ref{eq:GenVirTheorem} for a range of interaction strengths.
 In Fig. \ref{figure4} we plot the measured
difference $ T+I- V$ versus $(k_Fa)^{-1}$ along with $\frac{C}{4 \pi
k_F a}$, where we use our direct measurements of $C$. We find that
these independent measurements of the left and the right sides of
Eqn. \ref{eq:GenVirTheorem} agree to within our error, which is
roughly $1\%$ of the Fermi energy. It is interesting to note that
the measured energy difference $T+I-V$ is small (in units of $E_F$),
so that even a Fermi gas with a strongly attractive contact
interaction nearly obeys the non-interacting virial equation.

In conclusion, we have measured the integrated contact for a strongly interacting Fermi gas and demonstrated the connection between the
$1/k^4$ tail of the momentum distribution and the high frequency tail of rf spectra. Combining a measurement of $C$ vs $(k_Fa)^{-1}$ with measurements of the potential energy and the release energy of
the trapped gas, we verify two
universal relationships \cite{Tan08a,Tan08b}, namely the adiabatic
sweep theorem and the generalized virial theorem.  These universal
relations, which have now been experimentally confirmed, represent a
significant advance in the understanding of many-body quantum
systems with strong short-range interactions.  Furthermore, these
connections between microscopic and macroscopic quantities could be
exploited to develop novel experimental probes of the many-body
physics of strongly interacting quantum gases.

\begin{acknowledgments}
We acknowledge funding from the NSF and NIST.  We thank the JILA BEC group and also A. Perali and G. C. Strinati for helpful discussions.
\end{acknowledgments}

% You should use BibTeX and apsrev.bst for references
% Choosing a journal automatically selects the correct APS
% BibTeX style file (bst file), so only uncomment the line
% below if necessary.

\bibliographystyle{prsty}
%\bibliography{contact}

\end{document}